\documentclass[aps,prd,12pt,showpacs,preprint,groupedaddress,floatfix,nofootinbib]{revtex4}

\usepackage[fleqn]{amsmath}
\usepackage{amssymb}
\usepackage{graphicx}
\usepackage{color}
\usepackage{dcolumn}
\usepackage{bm}

\newcommand{\buno}{\mathbf{1}}
\newcommand{\bdos}{\mathbf{2}}
\newcommand{\btres}{\mathbf{3}}
\newcommand{\tp}{T^{\prime}}

\begin{document}

\preprint{DCP-10-05}

\title{A renormalizable fermion mass model with the double tetrahedral group}

\author{
 Alfredo Aranda,$^{1,2}$\footnote{Electronic address:fefo@ucol.mx}
 Cesar Bonilla,$^{1}$\footnote{Electronic address:rasec.cmbd@gmail.com}
 Raymundo Ramos,$^{1}$\footnote{Electronic address:raalraan@gmail.com}
 and Alma D. Rojas$^{1}$\footnote{Electronic address:alma.drp@gmail.com}}

\affiliation{$^1$Facultad de Ciencias, CUICBAS \\ 
Universidad de Colima, Colima, M\'exico \\
$^2$Dual C-P Institute of High Energy Physics, M\'exico \\}

\date{\today}

\begin{abstract}
We present a renormalizable model for fermion masses based solely on the double tetrahedral group $\tp$. 
It does not include right handed neutrinos and majorana neutrino masses are generated radiatively. The 
scalar sector of the model involves three SU(2) doublets and a set of lepton number violating (charged) scalars needed
to give mass to the neutrinos. In the quark sector the model leads to a Fritzsch type scenario that is consistent with
all the existing data. In the lepton sector, the model leads to tribimaximal (and near tribimaximal) mixing, and an inverted mass hierarchy.
\end{abstract}


\maketitle


\section{Introduction}
\label{sec:1}

Discrete symmetries have been extensively used in an attempt to describe the observed 
patterns in fermion masses and mixing angles (for a thorough list of references see~\cite{Altarelli:2010gt,Ishimori:2010au,Altarelli:2010fk}). Typical scenarios include some of the following ingredients:
\begin{itemize}
\item Introduction of a flavor symmetry, usually a product between a non-abelian and several abelian discrete groups. There can be different groups for different sectors (quark or leptonic) or only for one of the sectors.
\item Introduction of heavy right-handed neutrinos. This provides Dirac and Majorana mass terms for neutrinos and allows one to consider, for example, generation of small neutrino Majorana masses using some version of the seesaw mechanism~\cite{Schechter:1980gr}. 
\item Introduction of several scalar fields charged under the flavor symmetry that may or not acquire vacuum expectation values (vevs), thus breaking or not the flavor symmetry. These scalars may or not be singlets under the Standard Model (SM) gauge group. Typically there are high energy scales associated to the vevs of these fields (when they are singlets under the SM gauge group). With a few exceptions, the extended scalar sector can have collider relevant phenomenology~\cite{Tsumura:2009yf,Frampton:2010uw}.
\item Neutrino masses can be generated, in addition to the seesaw mechanism, through higher order operators that violate lepton number~\cite{Weinberg:1979sa} or through quantum (radiative) corrections~\cite{Zee:1980ai,Babu:1988qv,Fukugita:2003en}.
\end{itemize}

It is important to note that the ingredients above can be used in the frameworks (or combination thereof) of supersymmetric and non-supersymmetric models, Grand Unified Theories, extra dimensional models, and so on. Also the list is not exhaustive and includes only some of the most important and general considerations.

Among the possibilities generated by these ingredients, one that has been recently explored consists of creating renormalizable models~\cite{Frampton:2008ci,Frampton:2008bz,Morisi:2010rk}, i.e. models that forbid the inclusion of higher ($> 4$) dimension operators. There are two main challenges in this approach. First, it is highly nontrivial to create a model based on a single discrete group that can accommodate both quarks and leptons. Second, since no higher order operators are allowed, obtaining small neutrino masses is not trivial. For example in~\cite{Frampton:2008bz} a renormalizable model is investigated that uses the discrete group $\tp \times Z_2$, and where neutrino masses are generated by the seesaw mechanism. 
Another example is the work in~\cite{Morisi:2010rk} where a renormalizable model with flavor symmetry $G_F= S_4 \times Z_3^q\times Z_2^2 \rtimes (Z_{2e}\times Z_{2\mu} \times Z_{2\tau})$ is presented. Neutrino masses however are generated by the dimension 5, non renormalizable operator $LL\phi\phi/\Lambda$.

In this paper we present a model based on the single discrete group $\tp$, the double tetrahedral group, that accommodates both the quark and leptonic sectors. The model is renormalizable and neutrino masses are generated radiatively. In order to accomplish this we introduce several scalar fields charged under hypercharge and lepton number. $\tp$ is the smallest group we find to work under all of these considerations (see~\cite{Case:1956zz,Fairbairn:1964rr,Frampton:1994rk,Aranda:1999kc,Aranda:2000tm,Feruglio:2007uu,
Chen:2007afa,Frampton:2007et,Aranda:2007dp,Chen:2007gp,Sen:2007vx,Ding:2008rj,Chen:2009gy} for different models based on this group).

\section{A model with quarks, leptons and $\tp$}
\label{sec:2}

We present a model based on the discrete group $\tp$ that is renormalizable, i.e. we do not use higher dimension operators in order to generate patterns nor hierarchies. It only includes the Standard Model matter content and thus does not include right-handed neutrinos. Neutrino masses are thus zero at tree level and get generated only radiatively. The scalar sector is extended to include three SU(2) doublet fields, as well as a set of SU(2) singlet fields charged under lepton number and hypercharge that explicitly break lepton number conservation. These fields are needed in order to  generate neutrino masses radiatively.

$T'$ is the group of all 24 proper rotations in three dimensions leaving a regular tetrahedron invariant in the SU(2) double covering of SO(3)~\cite{Aranda:2000tm}. It has seven irreducible representations:\footnote{
See appendix \ref{a1} for more details about $T'$.} three singlets, $\mathbf{1^0}$ and $\mathbf{1^{\pm}}$, three doublets, $\mathbf{2^0}$ and
$\mathbf{2^{\pm}}$, and one triplet, $\mathbf{\mathbf{3}}$, with the following multiplication rules:
\begin{eqnarray}
    \begin{array}{ll}
      \mathbf{1}\otimes R=R\otimes \mathbf{1}=R & \mathbf{2}\otimes \mathbf{2}= \mathbf{1}\oplus \mathbf{3}\\
       \mathbf{2}\otimes\mathbf{3}=\mathbf{3}\otimes \mathbf{2}=\mathbf{2}^0\oplus \mathbf{2}^+ \oplus \mathbf{2}^- & \mathbf{3}\otimes\mathbf{3}=
       \mathbf{3}\oplus\mathbf{3}\oplus\mathbf{1}^0\oplus\mathbf{1}^+\oplus\mathbf{1}^- \\
    \end{array}
\end{eqnarray}
where $R$ stands for any representation.  Triality flips sign under Hermitian conjugation.

As stated above, the matter content of the model is exactly that of the SM. All matter fields transform as 
$\mathbf{2}^0\oplus\mathbf{1}^0$ under $\tp$, and there are no right-handed neutrinos. The scalar sector 
contains three SU(2) doublets ($Y=1/2$): $H^S$, and $H^D=(H_1^D,H_2^D)$ with the $\tp$ transformations
\begin{eqnarray}\label{H-transformations}
	H^S\sim \mathbf{1}^0, \ H^D \sim \mathbf{2},
\end{eqnarray}
and three SU(2) singlet, $Y=-1$, $L=2$ scalars $h^S$ and $h^D=(h_1^D, \ h_2^D)$ transforming as
\begin{eqnarray}\label{hs}
       h^S \sim\mathbf{1}^0 \ \ {\rm and} \ \ h^D \sim \mathbf{2}^0.
\end{eqnarray}

Given these assignments the Yukawa mass matrices have the following representation structure:
\begin{eqnarray}\label{yukawa-reps}
    Y_{u,d,l} \sim \left(\begin{array}{c|c}
\btres\oplus\buno^0 & \bdos^0\\
\hline \bdos^0 & \buno^0
\end{array}\right),
\end{eqnarray}
and all neutrinos are massless.

Fermion masses (except for neutrinos) are generated when the electroweak symmetry gets broken by the
vacuum expectation values (vevs) of the SU(2) doublet scalars, which we assume take the following form:\footnote{ This is an assumption. A complete analysis of the scalar sector is beyond the scope of this paper and will be presented
in a future publication.}
\begin{eqnarray}\label{Hvevs}
\langle H^D\rangle=(0,v_2),\:\:\mbox{and}\:\: \langle H^S\rangle=(v_s).
\end{eqnarray}

We now proceed to describe the quark and lepton sectors separately.

\subsection{Quark sector}
\label{subsec:quarks}

Given the $\tp$ representations for quarks and SU(2) doublets we have
\begin{eqnarray}
\mathcal{L}_{qY}&=&Y_S^u\bar{Q}_Du_{RD}\tilde{H}^S+Y_1^u\bar{Q}_D t_R\tilde{H}^D+Y_2^u\bar{t}_Lu_{RD}\tilde{H}^D+Y_S^t\bar{t}_Lt_R\tilde{H}^S+
\nonumber\\
&+&Y_S^d\bar{Q}_Dd_{RD}H^S+Y_1^d\bar{Q}_Db_{R}H^D+Y_2^d\bar{b}_Ld_{RD}H^D+Y_S^b\bar{b}_Lb_{R}H^S+ \ \rm{h.c.},
\end{eqnarray}
where $\tilde{H}=\imath \sigma_2 H^{*}$. Once the scalars acquire vevs we obtain the following mass matrices:
\begin{eqnarray} \label{mumdoriginal}
M_{u,d}=\left(\begin{array}{ccc}
0 & Y_S^{u,d}v_s& 0\\
-Y_S^{u,d}v_s & 0 & Y_1^{u,d}v_2\\
 0 & Y_2^{u,d}v_2 & Y_S^{t,b}v_s
 \end{array}
\right),
\end{eqnarray}
where each entry depends on both the values of the vevs and the Yukawa couplings. We see that these matrices can be parametrized as
\begin{eqnarray}\label{QMM}
M_{u,d} = \left( \begin{array}{ccc}
0 & A_{u,d} & 0 \\
-A_{u,d} & 0 & B_{u,d} \\
0 & D_{u,d} & C_{u,d}
\end{array} \right).
\end{eqnarray}

These matrices are similar to those found by Fritzsch for a three family model~\cite{Fritzsch:1977vd} (see also~\cite{Morisi:2010rk}). 
Following that analysis and taking $C_{u,d}=y^{2}_{u,d} m_{t,b}$, we rewrite the mass matrices above in terms of the quark masses and free parameters $y_{u,d}$~\cite{Babu:2004tn,Morisi:2010rk},
\begin{eqnarray}\label{mumdfinal}
\hat{M}_{u,d} =m_{t,b} \left( \begin{array}{ccc}
0 & q_{u,d}/y_{u,d} & 0 \\
-q_{a,d}/y_{u,d} & 0 & b_{u,d} \\
0 & d_{u,d} & y^{2}_{u,d} 
\end{array} \right) \ ,
\end{eqnarray}
where
\begin{eqnarray}
q^{2}_{u,d} = \frac{m_{u,d}m_{c,s}}{m_{t,b}^2},
\end{eqnarray}
\begin{eqnarray}
p_{u,d} = \frac{m_{u,d}^2 + m_{c,s}^2}{m_{t,b}^{2}},
\end{eqnarray}
\begin{eqnarray}
d_{u,d} =  \sqrt{\frac{p_{u,d} + 1 - y_{u,d}^{4} + R_{u,d}}{2} - \left(\frac{q_{u,d}}{y_{u,d}}\right)^2},
\end{eqnarray}
\begin{eqnarray}
b_{u,d} = \sqrt{\frac{p_{u,d} + 1 - y_{u,d}^{4} - R_{u,d}}{2} - \left(\frac{q_{u,d}}{y_{u,d}}\right)^2},
\end{eqnarray}
\begin{eqnarray}
R_{u,d} = ((1 + p_{u,d} - y_{u,d}^{4})^{2} - 4(p_{u,d} + q_{u,d}^{4}) + 8q_{u,d}^{2}y_{u,d}^{2})^{1/2},
\end{eqnarray}
and where $\hat{M}_{u,d}$ are matrices with real entries obtained from the phase factorization
of $M_{u,d}$~\cite{Babu:2004tn} through
\begin{eqnarray}
M_{u,d}=P_{u,d}\hat{M}P^{\ast}_{u,d}
\end{eqnarray}
with $P_{u,d}$ diagonal phase matrices such that
$P=P_{u}^{\dagger}P_{d}=\operatorname{diag}(1,e^{i\beta_{u,d}},e^{i\alpha_{u,d}})$. 

The free parameters are then $y_{u,d}$, $\alpha_{u,d}$, and $\beta_{u,d}$, and the CKM matrix is given by
\begin{eqnarray}
V_{CKM}=\mathcal{O}^{T}_{u}P\mathcal{O}_{d},
\end{eqnarray}
where the $\mathcal{O}_{u,d}$ matrices diagonalize $\hat{M}^{2}_{u,d}$ via,
\begin{eqnarray}
\mathcal{O}^{T}_{u,d}\hat{M}_{u,d}\hat{M}^{T}_{u,d}\mathcal{O}_{u,d}=\operatorname{diag}(m^{2}_{u,d},m^{2}_{c,s},m^{2}_{t,b}).
\end{eqnarray}

Using the values $y_u=0.9964$, $y_d=0.9623$, $\alpha_{u,d}=1.9560$, and $\beta_{u,d}=1.4675$ we obtain
\begin{eqnarray}
V^{th}_{CKM} = \mathcal{O}_u^{\dag}P \mathcal{O}_d=\left( \begin{array}{ccc}
0.974386 & 0.224853 & 0.00363\\
0.224723  & 0.973587 & 0.0403354 \\
0.00844 & 0.0396092  & 0.99918
\end{array} \right) \ ,
\end{eqnarray}
in perfect agreement with the experimental data~\cite{PDG2010}
\begin{eqnarray}
V_{CKM} =\left( \begin{array}{ccc}
0.97428\pm0.00015 & 0.2253 \pm 0.0007   & 0.00347 ^{+0.00016} _{-0.00012} \\
0.2252 \pm {0.0007}    & 0.97345^{+0.00015}_{-0.00016}  & 0.0410 ^{+0.0011}_{-0.0007} \\
0.00862^{+0.00026}_{-0.00020} & 0.0403 ^{+0.0010}_{-0.0007}  & 0.999152^{+0.000030}_{-0.000045}
\end{array} \right) \ .
\end{eqnarray}

\subsection{Lepton sector}
\label{subsec:leptons}

The analysis made for the quark sector extends directly to the charged leptons. The mass matrix can then be written
as before (see Eq.(\ref{mumdfinal})):
\begin{eqnarray}\label{mlfinal}
\hat{M}_{l} =m_{\tau} \left( \begin{array}{ccc}
0 & a_{l} & 0 \\
-a_{l}& 0 & b_{l} \\
0 & d_{l} & y^{2}_{l}
\end{array} \right)  .
\end{eqnarray}

Calling $U_l$ the matrix that diagonalizes $\hat{M}_l^2$, and using the values for the charged lepton masses taken from~\cite{PDG2010}, we obtain  
\begin{eqnarray}\label{Ul th}
   U_l= \left(%
\begin{array}{ccc}
 0.997042  & 0.0768522 & -0.000382008 \\
  0.0624654 &-0.813271  & -0.578522 \\
  -0.0447713 & 0.576787 &  -0.815667\\
\end{array}%
\right),
\end{eqnarray}
with $y_l = 0.9$ and $\alpha_l=\beta_l=0$.

For neutrinos the situation is different. We are assuming that neutrinos are Majorana type and that their masses can be induced by radiative corrections, thus making them light naturally~\cite{Babu:1988qv}.

In absence of right-handed neutrinos the only possible mass terms for left-handed neutrinos are Majorana mass terms. The simplest mass term in this case, without the introduction of scalars with non trivial SU(2) representations, is
the dimension five operator with form $\mathcal{L} \propto \bar{l}_L^c l_L \frac{HH}{M}$. Although this term is non-renormalizable, it may be induced by radiative corrections if we introduce a scalar field that breaks lepton 
number (provided there are at least two SU(2) Higgs doublets \cite{Zee:1980ai}). This is why we have introduced the fields $h^S$ and $h^D$ in our renormalizable model. 

In order to see how this works, consider the following example: A two Higgs doublet model with SM fermion content and an additional scalar field $h$ with charges $(1,-1)$ under $SU(2)\times U(1)_Y$ and lepton number $L=2$~\cite{Zee:1980ai,Fukugita:2003en}. The Yukawa couplings of $h$ are
\begin{eqnarray}\label{Lyuk}
    \mathcal{L}_{llh}=\kappa^{ab}\epsilon_{ij}\overline{(L^a_{i})^c}L^b_{j}h^{\dag}+
    h.c. \ ,
\end{eqnarray}
where $i,j$ are SU(2) indices, $a,b$ are family indices, $\kappa^{ab}=-\kappa^{ba}$ from Fermi statistics, and $L_i$ denotes the SU(2) lepton doublets. 
If there are two (or more) Higgs doublets, there will be a cubic coupling term like
\begin{eqnarray}\label{Lcubic}
    \mathcal{L}_{HHh}=\lambda_{\alpha\beta}\epsilon_{ij}H_i^{\alpha}H_j^{\beta}h + h.c.,
\end{eqnarray}
with $\lambda_{\alpha\beta}=-\lambda_{\beta\alpha}$, and $\alpha,\beta=1,2$. This term explicitly violates
lepton number and allows the generation of majorana masses for the neutrinos.

\begin{figure}[ht]
\includegraphics[width=10cm]{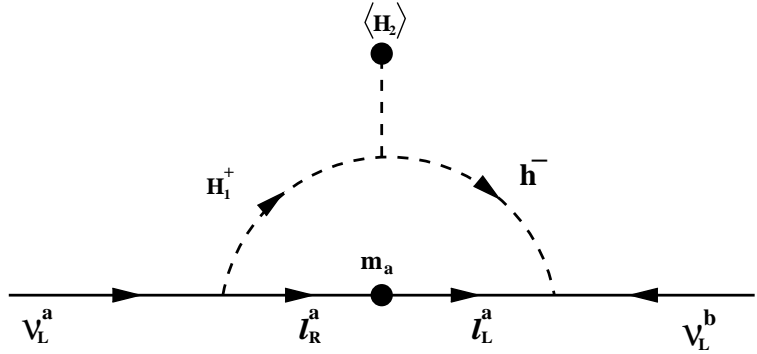}
\caption{One loop diagram giving rise to neutrino Majorana mass.}
\label{fig:loop}
\end{figure}

Notice that Eqs.~(\ref{Lyuk}) and (\ref{Lcubic}) (together with the usual Yukawa term from the lepton sector)
lead to the diagram shown in figure~\ref{fig:loop} that contributes
to a Majorana mass term as
\begin{eqnarray}\label{Mab}
M_{ab}=(-1)\kappa^{ab}m_a^2\frac{\lambda_{12}v_2}{v_1}\frac{1}{(4\pi)^2}\frac{1}{m_{H_1}^2-m_h^2
    }\log\frac{m_{H_1}^2}{m_h^2},
    \end{eqnarray}
where $m_{H_1}$ denotes the charged Higgs mass and $m_h$ the mass of the singlet 
field $h$.\footnote{We note that this is not yet in the scalar mass basis since the $H^0 H^{+}h^{-}$ term induces mixings between $H^{+}$ and $h^{-}$ However, we expect $m_H << m_h$, and work in the approximation that treats $m_h$ and $m_{H_1}$ as the physical masses.} Thus, the total contribution, including the diagram with $\nu^b_L$ and $\nu^a_L$ interchanged 
(which has the same form as in (\ref{Mab}) but with $a\leftrightarrow b$) is
\begin{eqnarray} \label{mab}\nonumber
m_{ab} &=& \kappa^{ab}(m_b^2-m_a^2)\frac{\lambda_{12}v_2}{v_1}\frac{1}{(4\pi)^2}\frac{1}{m_{H_1}^2-m_h^2
    }\log\frac{m_{H_1}^2}{m_h^2} \\
       &=& \kappa^{ab}(m_b^2-m_a^2)\frac{\lambda_{12}
    v_2}{v_1}F(m_h^2,m_{H}^2),	
    \end{eqnarray}
with~\cite{Fukugita:2003en}
\begin{eqnarray}\label{F}
    F(x,y)=\frac{1}{16\pi^2}\frac{1}{x-y}\log\frac{x}{y},
\end{eqnarray}

Coming back to our model, the terms relevant for neutrino mass generation are then given by
\begin{eqnarray}\label{lyuk}
\mathcal{L}_{Yuk } &=& Y_{1}\bar{L}^{D}\ell_{R}^D H^{S}+Y_{2}\bar{L}^{D}\tau_{R}H^{D}+Y_{3}\bar{L}_{\tau}\ell_{R}^D
H^{D}+Y_{\tau}\bar{L}_{\tau}\tau_{R}H^{S}+ \ h.c.\ , \\
\label{lLLh}
\mathcal{L}_{LLh}&=&\kappa^{mn} \overline{(L^{D}_m)^c} L^{D}_n h^S+\kappa^{m3} \overline{(L^{D}_m)^c} L_{\tau} h^D+\kappa^{3n} \overline{(L_{\tau})^c} L^{D}_n 
h^D+ \ h.c. \ , \\
 \label{lHHh} 
\mathcal{L}_{HHh} & = & \lambda_1 H^D H^S h^D \ + h.c. \ , 
\end{eqnarray}
where we have omitted SU(2) indices, $L = L^D \oplus L_{\tau}$ denote the left handed lepton SU(2)
doublets, and $\ell_R = \ell_R^D \oplus \tau_R$ denote the right handed lepton SU(2) singlets (recall
that the fermion fields transform as $\mathbf{2}^0\oplus\mathbf{1}^0$ under $\tp$).
Eq.(\ref{lyuk}) is the usual Yukawa term for leptons, and Eq.(\ref{lLLh}) comes 
from the general term in Eq.(\ref{Lyuk}), with $m,n=1,2,$ family indices.

Finally Eq.(\ref{lHHh}) contains the allowed trilinear coupling of $h^D$ with the SU(2) 
doublet scalar fields present in
the model. Eq.(\ref{lHHh}) represents the explicit lepton number violating term involving the scalar fields $h^S$ and $h^D$ necessary in order to generate majorana masses for the neutrinos.

The resulting non zero neutrino mass matrix elements are then given by
\begin{align}
    m_{\nu_e\nu_e}= & -\kappa^{13}m_{\tau\mu}Y_{e\mu}\lambda_1 v_2 F(m_H^2,m_h^2)
\end{align}
 \begin{align}
 m_{\nu_{e}\nu_{\mu}} = & \ m_{\nu_{\mu}\nu_{e}} =\left\{-\kappa^{23}m_{\tau\tau} Y_{e\tau}\lambda_1 v_s + \kappa^{13}m_{\tau\tau} Y_{\mu \tau}\lambda_1 v_s \right\} F(m_H^2,m_h^2)
\end{align}
\begin{align}
   m_{\nu_{e}\nu_{\tau}}  = & \ m_{\nu_{\tau}\nu_{e}}  = \left\{\kappa^{31}m_{e \mu} Y_{e\mu}\lambda_1 v_2-\kappa^{32}m_{\mu\tau} Y_{e\tau}\lambda_1 v_s
   +\kappa^{13}m_{\tau\mu} Y_{\tau\mu}\lambda_1 v_s \right.\nonumber \\
     & \left. + \kappa^{13}m_{\tau\tau} Y_{\tau \tau}\lambda_1 v_2\right\} F(m_H^2,m_h^2)
\end{align}
\begin{align}
   m_{\nu_{\tau}\nu_{\tau}} = &\left\{ -\kappa^{32}m_{\mu e} Y_{\tau e}\lambda_1 v_s +\kappa^{31}m_{e \mu} Y_{\tau \mu}\lambda_1 v_s \right\}F(m_H^2,m_h^2)
\end{align}

Assuming that $\kappa\sim$~O(1), $\lambda_1 \sim m_H\sim 500$~GeV, and noting that
$Y\langle H\rangle$ must be at the same scale of $m_l$, then $m_h\sim 4\times 10^5$~GeV
leads to matrix elements of O(eV). The Majorana neutrino mass matrix then has the texture:
\begin{eqnarray}\label{texture}
    M_{\nu}=\left(%
\begin{array}{ccc}
  a &b  & c \\
  b & 0 & 0 \\
  c & 0 & d \\
\end{array}%
\right),
\end{eqnarray}
where all entries are O(eV).

The neutrino mixing matrix $U_{MNS}$ is obtained by
\begin{eqnarray}\label{UMNS}
    U_{MNS}=U_l^{\dag} U_{\nu},
\end{eqnarray}
where $U_{l}$ is the unitary matrix that diagonalizes the square of the charged leptons matrix, and $U_{\nu}$ is the unitary matrix that diagonalizes $M_{\nu}$.

In order to perform the numerical analysis we used the following experimental results~\cite{PDG2010}:
\begin{eqnarray}\label{lepton-angles}
\sin^2(2\theta_{12}) & = & 0.087 \pm 0.03 \\
\sin^2(2\theta_{23}) & > & 0.92 \\
\sin^2(2\theta_{13}) & < & 0.15 
\end{eqnarray}
and
\begin{eqnarray}\label{lepton-masses}
\Delta m_{21}^2 & = & 7.59_{-0.21}^{+0.19}\times 10^{-5} \ \rm{eV}^2 \\
\Delta m_{32}^2 & = & 2.43 \pm 0.13\times 10^{-3} \ \rm{eV}^2 \ .
\end{eqnarray}
Since the absolute mass scale in the neutrino sector is not known, we use the following ratio
\begin{eqnarray}\label{mass-ratio}
0.0338 < \left| \frac{\Delta m_{21}^2}{\Delta m_{32}^2} \right| < 0.0288 \ .
\end{eqnarray}

\begin{figure}[ht]
\includegraphics[width=10cm]{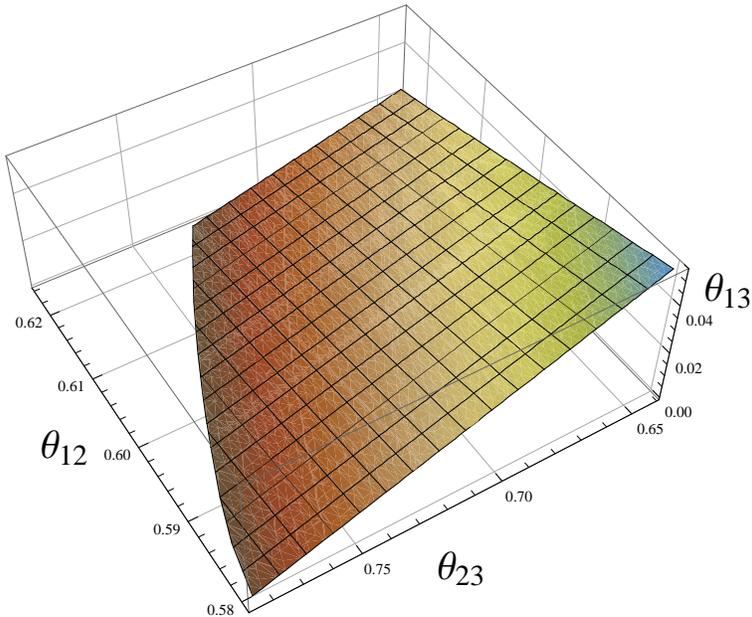}
\caption{Complete range for $\theta_{12}$, $\theta_{23}$, and $\theta_{13}$ (in radians) that can be obtained with the model presented in this paper. All points on this surface represent cases where the mass ratio defined in Eq.(\ref{mass-ratio}) falls in the desired range.}
\label{fig:angles}
\end{figure}

In order to determine whether the mass matrices in this model can reproduce these results, we performed a scan of the complete range in all three angles. Then for each case where a solution consistent with all three angles was found, we computed the ratio in Eq.(\ref{mass-ratio}) and selected those solutions that fell within its allowed range. We found that solutions exist with the following properties:

\begin{itemize}
\item All solutions have an inverted hierarchy for neutrino masses.
\item There are solutions for all the allowed ranges in $\theta_{23}$ and $\theta_{12}$ but 
only for $\sin^2(2\theta_{13}) < 0.012$ (see figure~\ref{fig:angles}).
\end{itemize}

\section{Conclusions}
\label{sec:4}
We sought to find a renormalizable fermion mass model based on a single discrete group, and with the minimum number of additions to the Standard Model. We found that it is possible to do so provided that neutrino masses are generated radiatively, and that no right handed neutrinos are present. The smallest discrete group that worked under these considerations is the double tetrahedral group $\tp$.

A specific realization of a model is presented. It contains the Standard Model fermion content and a scalar sector that involves three SU(2) doublets as well as a set of lepton number violating (charged) scalars, needed to give mass to the neutrinos. 

In the quark sector the model leads to a Fritzsch type scenario that is consistent with
all the existing data. In the lepton sector, the model leads to tribimaximal (and near tribimaximal) mixing and an inverted mass hierarchy: it accommodates all existing data in both the quark and lepton sectors with
$\sin^{2}(2\theta_{13}) < 0.012$. The phenomenology of the scalar sector is currently being studied and will be presented in a future publication.

\acknowledgments
This work was supported in part by CONACYT.

\appendix
\section{Useful information about $\tp$}\label{a1}

The products of $\tp$ representations are given by
\begin{eqnarray} 
&&{\bf 1} \otimes {\bf \mathcal{R}} = {\bf \mathcal{R}} \otimes {\bf 1} = \mathcal{R} \ \forall \ {\rm irrep} \mathcal{R}
\\ 
&&{\bf 2}^{0,\pm} \otimes {\bf 3} = {\bf 3} \otimes {\bf 2}^{0,\pm} = {\bf 2^{0}}\oplus {\bf2^{+}}\oplus {\bf 2^{-}}
\\
&&{\bf 3} \otimes {\bf 3} = {\bf 1^{0}}\oplus {\bf1^{+}}\oplus {\bf 1^{-}}\oplus{\bf3}\oplus{\bf3}
\\
&&{\bf 2^{0,\pm}} \otimes {\bf 2^{0,\mp}} = {\bf 3} \oplus {\bf 1}.
\end{eqnarray}

The rules for products between trialities are 
$\{\pm\}\otimes\{\pm\}=\{\mp\}$, $\{\mp\}\otimes\{\pm\}=\{0\}$, and 
$\{0\}\otimes\{0\}=\{0\}$. Under charge conjugation, trialities change sign.

The Clebsch-Gordan coefficients are given by:

\begin{eqnarray}
{\bf 1} &\otimes& {\bf 2}= {\bf 2} \sim\
\left(\begin{array}{c}
\alpha\beta_{1} \\
\alpha\beta_{2}
\end{array} \right)
\\
{\bf 1^{0;+;-}} &\otimes& {\bf 3}= {\bf 3} \sim\ \left(\begin{array}{c}
\alpha\beta_{1;3;2} \\
\alpha\beta_{2;1;3} \\ 
\alpha\beta_{3;2;1}
\end{array} \right)
\end{eqnarray}
\begin{eqnarray}
{\bf 2}^{0} &\otimes& {\bf 2}^{0}={\bf 2}^{\pm} \otimes {\bf 2}^{\mp}= {\bf1}^{0}\oplus{\bf3}\,\, \text{with}
\begin{cases}
{\bf 1}\sim -\alpha_{2}\beta_{1}+\alpha_{1}\beta_{2} &\\
{\bf 3}\sim\left(\begin{array}{c}
\dfrac{1}{2}\left(1-i\right)\left(\alpha_{1}\beta_{2}+\alpha_{2}\beta_{1}\right) \\
i\alpha_{1}\beta_{1} \\
\alpha_{2}\beta_{2}
\end{array} \right)
\end{cases}
\\
{\bf 2}^{0} &\otimes& {\bf 2}^{+}={\bf 2}^{-} \otimes {\bf 2}^{-}= {\bf1}^{+}\oplus{\bf3}\,\, \text{with}
\begin{cases}
{\bf 1}^{+}\sim -\alpha_{2}\beta_{1}+\alpha_{1}\beta_{2} &\\
{\bf 3}\sim\left(\begin{array}{c}
\alpha_{2}\beta_{2}\\
\dfrac{1}{2}\left(1-i\right)\left(\alpha_{1}\beta_{2}+\alpha_{2}\beta_{1}\right) \\
i\alpha_{1}\beta_{1}
\end{array} \right)
\end{cases}
\end{eqnarray}
\begin{eqnarray}
{\bf 2}^{0} &\otimes& {\bf 2}^{-}={\bf 2}^{+} \otimes {\bf 2}^{+}= {\bf1}^{-}\oplus{\bf3}\,\, \text{with}
\begin{cases}
{\bf 1}^{-}\sim -\alpha_{2}\beta_{1}+\alpha_{1}\beta_{2}  &\\
{\bf 3}\sim\left(\begin{array}{c}
i\alpha_{1}\beta_{1}\\
\alpha_{2}\beta_{2}\\
\dfrac{1}{2}\left(1-i\right)\left(\alpha_{1}\beta_{2}+\alpha_{2}\beta_{1}\right)
\end{array} \right)
\end{cases}
\\
{\bf 2}^{0;+;-} &\otimes& {\bf 3}={\bf 2}^{0} \oplus {\bf 2}^{+}\oplus {\bf 2}^{-}\,\, \text{with}
\begin{cases}
{\bf 2}^{0}\sim\left(\begin{array}{c}
\alpha_{1;1;1}\beta_{1;3;2}+(1+i)\alpha_{2;2;3}\beta_{2;1;2}\\
-\alpha_{2;2;2}\beta_{1;3;2}+(1-i)\alpha_{1;1;1}\beta_{3;2;1}
\end{array} \right) &\\
{\bf 2}^{+}\sim \left(\begin{array}{c}
\alpha_{1;1;1}\beta_{2;1;3}+(1+i)\alpha_{3;2;2}\beta_{2;2;1}\\
-\alpha_{2;2;2}\beta_{2;1;3}+(1-i)\alpha_{1;1;1}\beta_{1;3;2}
\end{array} \right) &\\
{\bf 2}^{-}\sim \left(\begin{array}{c}
\alpha_{1;1;1}\beta_{3;2;1}+(1+i)\alpha_{2;3;2}\beta_{1;2;2}\\
-\alpha_{2;2;2}\beta_{3;2;1}+(1-i)\alpha_{1;1;1}\beta_{2;1;3}
\end{array} \right)
\end{cases}
\end{eqnarray}
\begin{eqnarray}
{\bf 3} \otimes {\bf 3}={\bf 1}^{0} \oplus {\bf 1}^{+}\oplus {\bf 1}^{-}\oplus {\bf 3}_{s}\oplus {\bf 3}_{a}\,\, \text{with}
\begin{cases}
{\bf 1}^{0}\sim \alpha_{1}\beta_{1}+\alpha_{2}\beta_{3}+\alpha_{3}\beta_{2} &\\
{\bf 1}^{+}\sim \alpha_{1}\beta_{2}+\alpha_{2}\beta_{1}+\alpha_{3}\beta_{3} &\\
{\bf 1}^{-}\sim \alpha_{1}\beta_{3}+\alpha_{2}\beta_{2}+\alpha_{3}\beta_{1} &\\
{\bf 3}_{s}\sim \left(\begin{array}{c}
2\alpha_{1}\beta_{1}-\alpha_{3}\beta_{2}-\alpha_{2}\beta_{3}\\
2\alpha_{3}\beta_{3}-\alpha_{1}\beta_{2}-\alpha_{2}\beta_{1}\\
2\alpha_{2}\beta_{2}-\alpha_{1}\beta_{3}-\alpha_{3}\beta_{1}
\end{array} \right) &\\
{\bf 3}_{a}\sim \left(\begin{array}{c}
-\alpha_{2}\beta_{3}+\alpha_{3}\beta_{2}\\
-\alpha_{1}\beta_{2}+\alpha_{2}\beta_{1}\\
-\alpha_{1}\beta_{3}+\alpha_{3}\beta_{1}
\end{array} \right)
\end{cases}
\end{eqnarray}
where $\alpha_i$ and $\beta_i$ denote the entries in the first and second representations respectively.

\end{document}